\documentclass[prl,twocolumn]{revtex4}
\usepackage{graphicx}
\usepackage{hyperref}
\usepackage[section]{placeins}
\usepackage{amsmath}
\usepackage{amssymb}
\usepackage{dsfont}
\usepackage{bbold}
\usepackage{ulem}
\newcommand{\ket}[1]{|#1\rangle}             
\newcommand{\bra}[1]{\langle#1|}             

\begin{document}

\title{Entanglement verification via nonlinear witnesses}

\author{Megan Agnew,$^{1}$ Jeff Z.~Salvail,$^{1}$ Jonathan Leach,$^{1}$ Robert W.~Boyd$^{1, 2}$}
\affiliation{$^1$Dept.~of Physics, University of Ottawa, 150 Louis Pasteur, Ottawa, Ontario, K1N 6N5 Canada}
\affiliation{$^2$Institute of Optics, University of Rochester, Rochester, USA}
\date{\today}

\begin{abstract}

The controlled generation of entangled states and their subsequent detection are integral aspects of quantum information science. In this work, we analyse the application of nonlinear witnesses to the verification of entanglement, and we demonstrate experimentally that nonlinear witnesses perform significantly better than linear witnesses.   Specifically, we demonstrate that a single nonlinear entanglement witness is able to determine to a high degree of certainty that a mixed state containing orbital angular momentum (OAM) entanglement of the form $(\ket{j,j} + e^{i \varphi} \ket{k,k})/\sqrt{2}$ (or $(\ket{j,k} + e^{i \varphi} \ket{k,j})/\sqrt{2}$) is entangled for any relative phase $\varphi$ and sufficient fidelity; $j$ and $k$ are OAM azimuthal quantum numbers. This is a significant improvement over linear witnesses, which cannot provide the same level of performance. We envisage that nonlinear witnesses and our method of state preparation will have further uses in areas of quantum science such as superdense coding and quantum key distribution.

\end{abstract}
\maketitle

{\it Introduction}: Entanglement, which can produce nonlocal correlations that are stronger than those predicted by classical physics, is an essential part of quantum mechanics \cite{Reid2009}. As a result, it has been studied extensively as a means to test quantum mechanics \cite{EPR,Bell,Hardy1993}. Entanglement is a vital resource in many quantum information protocols. For example, entangled photonic qubits are important for communication protocols such as quantum key distribution \cite{Ekert1991,Gisin2002}, superdense coding \cite{Bennett1992,Barreiro2008}, and quantum teleportation \cite{Bennett1993,Bouwmeester1997,Marcikic2003}, and entangled qubits are required for implementations of quantum computing.  Consequently, the efficient detection of entangled states \cite{Guhne2009} plays a vital role in many quantum information science applications.

Determining the full quantum state of a system can be accomplished through tomography \cite{James2001,Resch2005,Agnew2011}, which consists of taking many measurements on identical copies of a quantum state. The resulting real-valued probabilities are then used to estimate the complex-valued state that best fits the measurements. Tomography can determine the complete density matrix that describes the system; however, it is inefficient for determining entanglement as it requires a large number of measurements. This is particularly relevant when considering high-dimensional or multipartite systems.  As the number of elements required to describe the state increases, so too does the total number of measurements required for reconstruction \cite{Thew2002} \footnote{For $n$ identical particles in $d$ dimensions, the number of measurements for a tomographically complete set scales as $d^{2n}-1$.}.  Compressive sensing is one approach to reducing the required number of measurements, but reconstructing the state from less information leads to a less accurate estimation of the final state \cite{Gross2010}.

Entanglement witnesses provide an alternative to tomography in the case where complete knowledge of the state is not required and detecting entanglement is the goal. An entanglement witness establishes directly whether a quantum state belonging to a certain class is entangled \cite{Lewenstein2000,Guhne2002,D'ArianoWitness2003,Pittenger2003,Bertlmann2005}. The use of entanglement witnesses can be more efficient than tomography as witnesses require fewer measurements and no reconstruction.  More generally, the application of witnesses in quantum science plays a vital role in establishing particular properties of systems \cite{Hendrych2012}.   

Linear entanglement witnesses, which are witnesses that depend linearly on expectation value, have been used to detect entanglement in bipartite polarisation states \cite{Barbieri2003} and orbital angular momentum (OAM) states \cite{Leach2010,Agnew2012}. Multipartite entanglement has also been detected using a linear entanglement witness \cite{Bourennane2004}. Linear entanglement witnesses are efficient as they require the fewest possible number of measurements that will give sufficient information about the state; however, in order to function optimally, they require prior knowledge of the form of the entangled state.  For example, different linear witness are required for each of the four Bell states $\ket{\Phi^{_+}}, \ket{\Phi^{_-}}, \ket{\Psi^{_+}}$ and $\ket{\Psi^{_-}}$; using a linear witness that is not appropriate to the form of the state may produce an inconclusive result when used to detect entanglement.

Alternatively, nonlinear entanglement witnesses improve upon an existing linear entanglement witness with a term that relies nonlinearly on expectation value \cite{Uffink2002,Giovannetti2003,Hyllus2006,Guhne2006}. The improvement is that a nonlinear witness is able to verify entanglement over a significantly larger set of states compared to its linear counterpart.  Returning to the example of the Bell states, one can construct a single nonlinear witness that will detect both correlated Bell states $\ket{\Phi^{_+}}$ and $\ket{\Phi^{_-}}$ and a single nonlinear witness that will detect both anti-correlated Bell states $\ket{\Psi^{_+}}$ and $\ket{\Psi^{_-}}$. Specifically, one nonlinear witness works for almost all correlated states regardless of relative phase between the modes, and one nonlinear witness works for almost all anti-correlated states. Importantly, these nonlinear witnesses are accessible when the nonlinear extension can be achieved using the same measurements as for the linear witness \cite{Arrazola2012}.
 
In this work, we demonstrate the controlled generation of a wide range of spatially entangled states and the subsequent experimental realisation of a class of nonlinear entanglement witnesses.   We compare the expectation values of the nonlinear witnesses to those of the standard linear witnesses and establish that the nonlinear witnesses are capable of detecting entanglement over a wide range of states.  The particular degree of freedom we choose to investigate is orbital angular momentum; however, our results are general in that the procedures can be applied to other degrees of freedom such as polarisation or spin. 
 
 \begin{figure}
 \centering
 \includegraphics[]{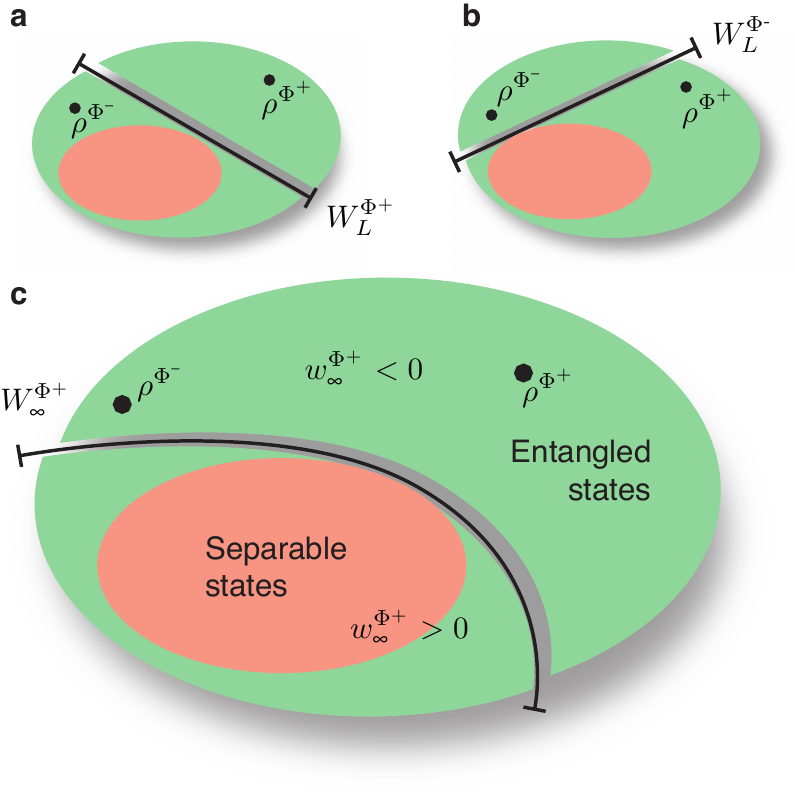}
 \caption{\footnotesize{{\bf Visual representation of nonlinear and linear entanglement witnesses.} The state $\rho^{\Phi^+}$ is entangled and is separated from the set of separable states by the linear entanglement witness $W_L^{\Phi^{_+}}$ ({\bf a}), and the entangled state $\rho^{\Phi^-}$ is separated from the set of separable states by the linear entanglement witness $W_L^{\Phi^{_-}}$ ({\bf b}). However, both states are separated from the set of separable states by the nonlinear witness $W_\infty^{\Phi^{_+}}$ ({\bf c}).}}
 \label{figNLEW}
\end{figure}

\vspace{0.5cm} 
{\it Theory}: The characteristic feature of entanglement is the observation of non-local correlations between two qubits that belong to spatially separated systems.  We denote the eigenstates of the qubits by $\ket{j}$ and $\ket{k}$ and the two spatially separated systems by $A$ and $B$. Then the general entangled state that contains two qubits in the same state, i.e. correlated qubits, can be written as 
\begin{equation}\label{ent_corr}
\ket{\Phi}=\frac{1}{\sqrt{1+\varepsilon^2}} \Bigl( \ket{j, j}+\varepsilon e^{i\varphi}\ket{k, k}\Bigr).
\end{equation}
Here, $\varepsilon$ defines the degree of entanglement, $\varphi$ is the phase between the modes and defines the nature of the correlations, and we use $\ket{j, j}$ to be equivalent to $\ket{j}_A \otimes \ket{j}_B$.  Two of the four Bell states $\ket{\Phi^\pm}$ are particular cases of equation (\ref{ent_corr}) where $\varepsilon$ is equal to unity such that the state is maximally entangled, and the phase $\varphi$ is equal to either $0$ ($\ket{\Phi^+}$) or $\pi$ ($\ket{\Phi^-}$). The general entangled state $\ket{\Psi}$ that contains two qubits in opposite states, i.e. anti-correlated qubits, can be denoted by replacing $\ket{j, j}$ with $\ket{j,k}$ and $\ket{k, k}$ with $\ket{k, j}$.  The remaining two Bell states $\ket{\Psi^\pm}$ are particular cases of the anti-correlated entangled state.

However, in practice, the incident state need not be pure; that is, the photons are described by
\begin{equation}
\rho^{\psi}=\ket{\psi}\bra{\psi} p + \mathbb{1}(1-p)/4,
\end{equation}
where $p$ is the probability of obtaining the entangled state $\ket{\psi}$ and $\mathbb{1}$ is the identity matrix, which represents uncoloured noise.  We use the superscript $\psi$ to indicate that the convex combination $\rho$ is partially composed of the entangled state $\ket{\psi}$.  Whether or not the state $\rho^{\psi}$ is entangled is determined by the probability $p$: states with $p>1/3$ are entangled \cite{Guhne2009}.

The use of entanglement witnesses is an efficient method to detect the entanglement of a state.  The expectation value $w$ of a witness $W$ on a quantum state $\rho$ provides the relevant information: a negative expectation value indicates entanglement, whereas a positive expectation value gives an inconclusive result.  If a positive expectation value is obtained, the information gained is that either the state is separable, or the witness chosen was not appropriate for the form of the entangled state. The simplicity of such a result has led to the widespread use of linear entanglement witnesses.

Linear entanglement witnesses are the simplest form of entanglement witness. However, linear witnesses function only over restricted sets of states: entanglement of the set of states $\rho^{\Phi}$ (or $\rho^{\Psi}$) cannot be verified with a single linear witness.  As an example, the entanglement of the Bell state $\ket{\Phi^{_-}}$ cannot be confirmed using the linear witness $W_L^{\Phi^{_+}}$ (constructed for the state $\ket{\Phi^{_+}}$) because the expectation value $w^{\Phi^{_+}}_L$ is positive.

Recently, it was shown that it is possible to improve a linear witness with a term that relies nonlinearly on expectation value \cite{Guhne2006,Moroder2008,Arrazola2012}. One improvement is that entanglement of a significantly larger fraction of the set of states $\rho^{\Phi}$ (or $\rho^{\Psi}$) can be verified with a single nonlinear witness that contains the same observables as the linear witness. For any value of $p$, there exists a nonlinear improvement of a linear witness that always verifies the entanglement of a larger set of states compared to its linear counterpart.  As an example, the entanglement of both the Bell states $\ket{\Phi^{_-}}$ and $\ket{\Phi^{_+}}$ can be confirmed using a single nonlinear witness. A visual comparison between linear and nonlinear witnesses is shown in figure~\ref{figNLEW}. 

\begin{figure*}
  \centering
  \includegraphics[]{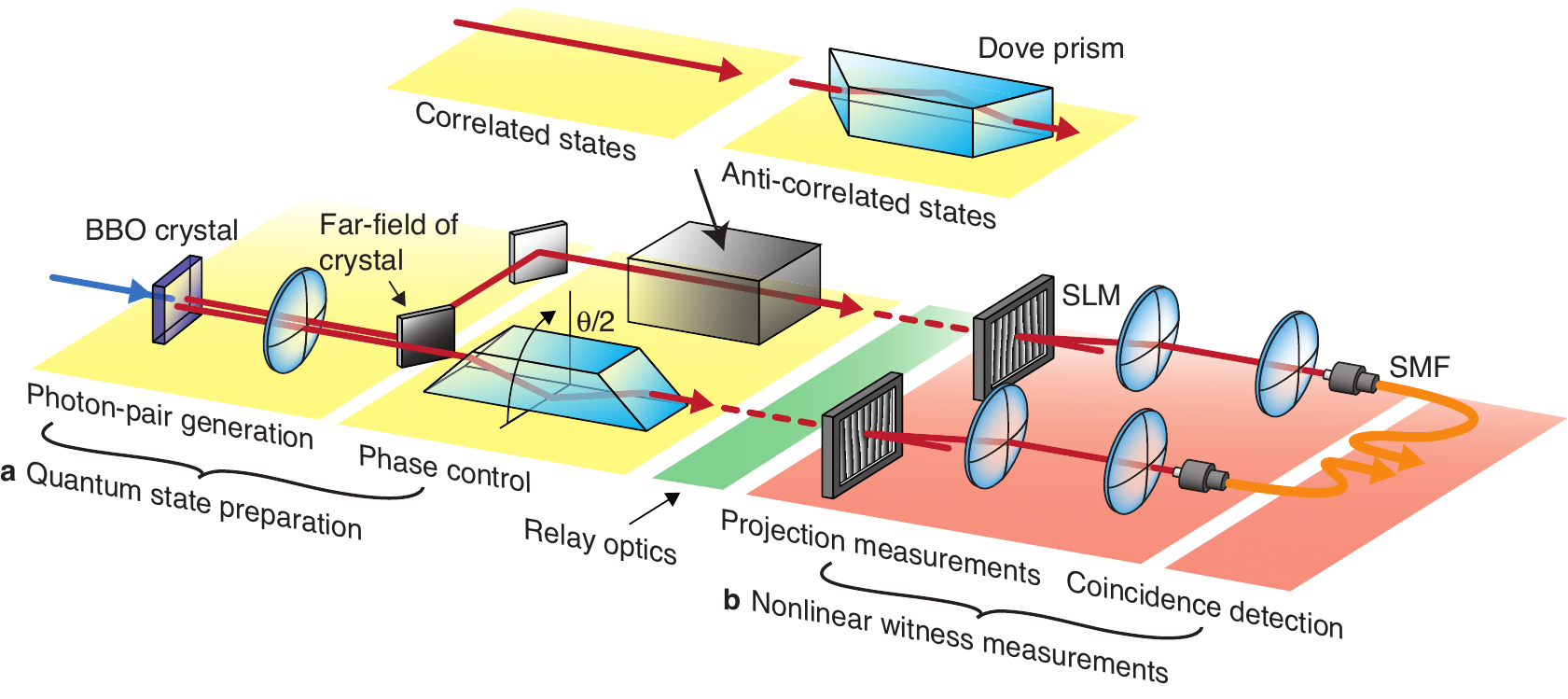} 
  \caption{\footnotesize{{\bf Schematic of the setup}. The experiment has two main stages:  ({\bf a}) quantum state preparation and  ({\bf b}) nonlinear witness measurements.  In the quantum state preparation stage, entangled photon pairs are generated by parametric downconversion.  The number of Dove prisms then allows us to prepare either a correlated (one prism) or anti-correlated state (two prisms), and the angle of the first prism gives control of the phase between the constituent modes.   For the nonlinear witness measurements,  projective measurements are made with spatial light modulators (SLMs) and single-mode fibres (SMFs) used in combination with single-photon detectors and coincidence detection electronics.}}
  \label{setupnl}
\end{figure*}

In our experiment, we compare the accessible nonlinear witnesses $W^{\Phi^{_+}}_\infty$ and $W^{\Psi^{_+}}_\infty$ to their corresponding linear witnesses. In order to construct the nonlinear improvement to the linear witnesses, we use the method outlined in Ref.~\cite{Arrazola2012}. One starts with the original linear witness $W^{\Phi^{_+}}_L$, which can be considered a first-order witness $W^{\Phi^{_+}}_1$. The $n^{\rm th}$-order witness $W^{\Phi^{_+}}_n$ can be found by iteration \cite{Moroder2008,Arrazola2012}. Taking the limit as $n \rightarrow \infty$, one obtains $W_\infty^{\Phi^{_+}}$.

The expectation value $w_\infty^{\Phi^{_+}}$ of this witness can be expressed as a combination of expectation values of measurable operators.  Contained within the measurements for the nonlinear witness is a unitary operator $U$, which provides some freedom in choosing the exact form of the witness.   By choosing $U$ to be equal to $-\sigma_z \otimes \sigma_z$, we show in the supplementary information that the expectation value of the particular nonlinear witness that we consider in this experiment is given by
\begin{align}
w^{\Phi^{_+}}_\infty(\rho)&= {\rm Tr}(\rho W^{\Phi^{_+}}_L)-|{\rm Tr}(\rho W^{\Phi^{_+}}_L)|^2 \label{nlw}\\
&-\frac{|{\rm Tr}(\rho W^{\Phi^{_+}}_L)-{\rm Tr}(\rho W^{\Phi^{_+}}_L ){\rm Tr}(\rho (-\sigma_z \otimes \sigma_z))|^2}{1-|{\rm Tr}(\rho (-\sigma_z \otimes \sigma_z))|^2} \nonumber. 
\end{align}
We see that ${\rm Tr}(\rho W^{\Phi^{_+}}_L)$ and ${\rm Tr}(\rho (-\sigma_z \otimes \sigma_z))$ are the only measurements that are required for the nonlinear witness. A similar method may be used to generate the nonlinear improvement of the linear witness $W_L^{\Psi^{_+}}$, but in this case we require ${\rm Tr}(\rho (\sigma_z \otimes \sigma_z))$ to achieve the same result.  The general form of the nonlinear witness is given in the supplementary information.


For the correlated Bell states $\ket{\Phi^{_\pm}}$, the decomposition of the operator $W_L^{\Phi^{_\pm}}$ with the fewest local measurements is given by \cite{Guhne2002}
\begin{align}
W_L^{\Phi^{_\pm}}&=\frac{1}{2}  \Bigl( \ket{j, k}\bra{j, k}+\ket{k, j}\bra{k, j} \Bigr) \notag\\
&\pm \frac{1}{2} \Bigl( \ket{x^{_+},x^{_-}}\bra{x^{_+},x^{_-}}+\ket{x^{_-},x^{_+}}\bra{x^{_-},x^{_+}}\notag\\
&-\ket{y^{_+},y^{_-}}\bra{y^{_+},y^{_-}}-\ket{y^{_-},y^{_+}}\bra{y^{_-},y^{_+}}   \Bigr),\label{decompositioncorr}
\end{align}
and the unitary operator is 
\begin{align}
-&\sigma_z \otimes \sigma_z  \notag \\
& = \ket{j ,j}\bra{j, j}+\ket{k, k}\bra{k, k} - \underbrace{\left( \ket{j, k}\bra{j, k}+\ket{k, j}\bra{k, j} \right)}_\text{Contained within $W_L^{\Phi^{_\pm}}$} .
\end{align}
The linear witnesses $W_L^{\Psi^{_\pm}}$ have similar decompositions, and in all cases, 
\begin{align*}
\ket{x^{_\pm}} =\frac{1}{\sqrt{2}} \Bigl( \ket{j} \pm \ket{k} \Bigr) \text{ and } \ket{y^{_\pm}}& = \frac{1}{\sqrt{2}}  \Bigl( \ket{j} \pm i \ket{k} \Bigr).
\end{align*}
It follows that the nonlinear witness that we choose to investigate has a total of eight projective measurements, which is a nearly twofold improvement over the number required for complete tomography of the state \cite{Thew2002}.  This is because there are always two measurements that occur in both $W_L^{\Phi^{_\pm}}$ and $-\sigma_z \otimes \sigma_z $ or in both $W_L^{\Psi^{_\pm}}$ and $\sigma_z \otimes \sigma_z $.  For the correlated case these are $\ket{j,k}\bra{j, k}$ and $\ket{k, j}\bra{k, j} $, and for the anti-correlated case these are $\ket{j,j}\bra{j, j}$ and $\ket{k, k}\bra{k, k}$.  Additionally, we note that ${\rm Tr}(\rho (-\sigma_z \otimes \sigma_z))$ is the measure of the strength of the correlations.  Perfect correlations (or anti-correlations) correspond to $|{\rm Tr}(\rho (-\sigma_z \otimes \sigma_z))|$  equal to unity, and no correlations correspond to $|{\rm Tr}(\rho (-\sigma_z \otimes \sigma_z))|$ equal to zero. 

\vspace{0.5cm} 
{\it Quantum state preparation}: To experimentally realise nonlinear entanglement witnesses, we require precise control of the form of the entangled state; see figure \ref{setupnl}.  To fully test the nonlinear witnesses, we need to prepare a range of correlated $\rho^{\Phi}$ and anti-correlated $\rho^{\Psi}$ states.  As the benefit of the particular nonlinear witnesses that we are investigating is that they detect entangled states regardless of the phase between the modes, we require the ability to adjust this parameter.

Our investigation concerns entanglement of the spatial degree of freedom.  More specifically, we look for entanglement between orbital angular momentum states of light in two-dimensional state spaces.  Consequently, we can achieve exact quantum state preparation through the use of unitary transformations applied to the signal and idler photons of the downconverted light. We accomplish such transformations using Dove prisms; the number of prisms allows us to choose between a correlated and anti-correlated entangled state, and the angle of the prisms allows us to manipulate the phase between the entangled modes.

A Dove prism placed at an angle $\theta/2$ performs two actions on the transmitted light: firstly, the transverse cross-section of any transmitted beam is reversed such that $\ell \rightarrow -\ell$; secondly, an $\ell$-dependent phase shift is introduced such that the modes within the beam acquire the additional phase $\ell \theta$.   It follows that two Dove prisms can be used to introduce an $\ell$-dependent phase shift between different OAM modes whilst leaving the sign of $\ell$ unchanged.

Placing a Dove prism oriented at an angle $\theta_B/2$ in arm B of the downconversion system, we obtain a correlated entangled state.  Of particular relevance are the two-dimensional subspaces that include the $\ket{\ell =0}$ mode as these are correlated entangled states of the form of equation (\ref{ent_corr})
\begin{align}\label{6}
\ket{\Phi_\ell}=\frac{1}{\sqrt{1+\varepsilon_\ell^2}} \Bigl( \ket{0, 0}+\varepsilon_\ell e^{i \varphi}\ket{\ell, \ell} \Bigr),
\end{align}
where $\varphi = \ell \theta_B$. Placing a second Dove prism oriented at an angle $\theta_A/2$ in arm A of the system converts the state $\ket{\Phi_\ell}$ to an anti-correlated state
\begin{align}\label{7}
\ket{\Psi_\ell}=\frac{1}{\sqrt{1+\varepsilon_\ell^2}} \Bigl( \ket{0, 0}+\varepsilon_\ell e^{i \phi}\ket{-\ell, \ell} \Bigr),
\end{align}
where $\phi = \ell (\theta_B- \theta_A)$.  Using the states $\ket{\Phi_\ell}$ and $\ket{\Psi_\ell}$, we can test the ability of the nonlinear witness for detecting entanglement for a large range of different states. Although we do not use them, we also note that all four Bell states can be produced by this method of quantum state preparation.

\vspace{0.5cm}
{\it Experiment results}: Before we calculate any expectation value, we perform quantum state tomography on each input state to ensure that it is indeed entangled.  We find that our method of quantum state preparation is able to produce the desired quantum states of the form given in equations (\ref{6}) and (\ref{7}) to a high degree of confidence.  Thus, we then proceed to calculate the relevant expectation values so that we can assess the performance of nonlinear and linear witnesses for the verification of entanglement. 
 \begin{figure}[h!]
  \centering
  \includegraphics[]{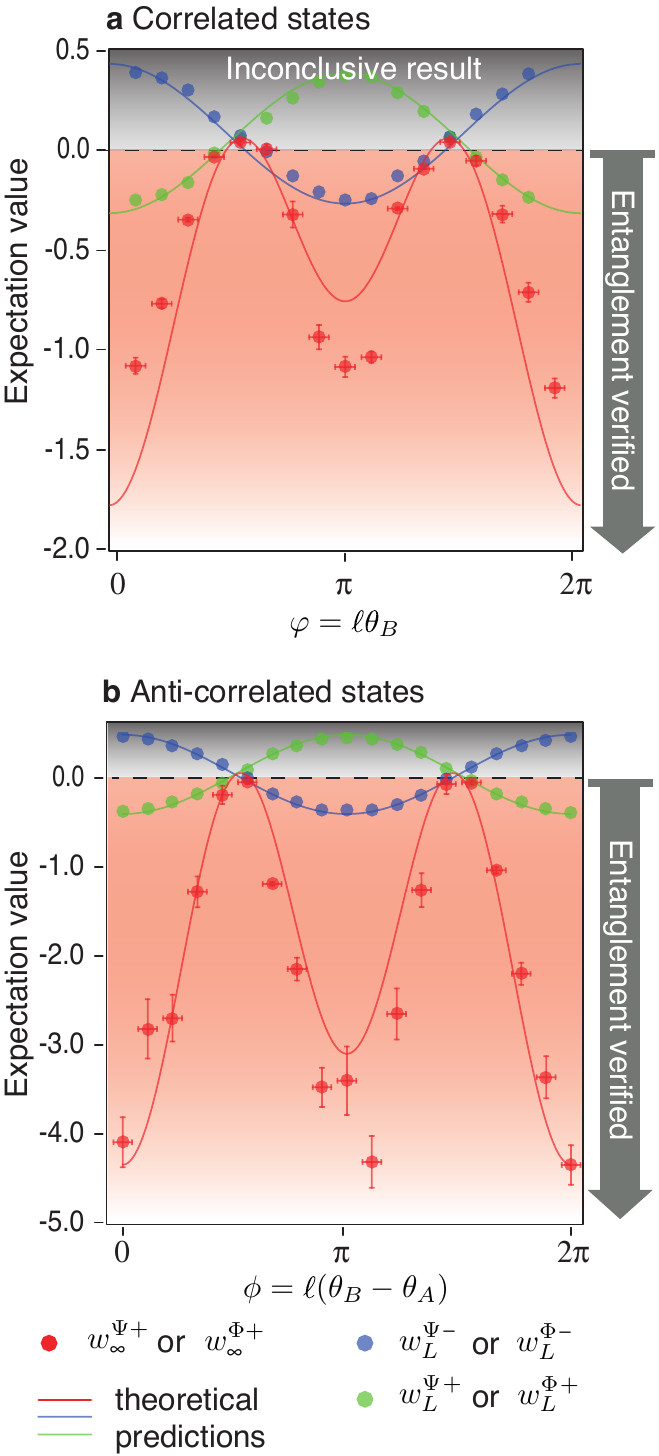} 
  \caption{\footnotesize{{\bf Expectation values for nonlinear witnesses and linear witnesses.} Experimentally recorded expectation values of the nonlinear entanglement witness and the two linear witnesses for correlated ({\bf a}) and anti-correlated ({\bf b}) entangled states for $\ell = 2$. A negative expectation value indicates that the state is entangled.  Note that the nonlinear witness correctly detects entanglement under almost all cases, whereas each of the linear witnesses often fails to detect entanglement, even though it is present.  For the correlated states, we measured $w_\infty^{\Phi^{_+}}$, $w_L^{\Phi^{_+}}$, and $w_L^{\Phi^{_-}}$, and for the anti-correlated states, we measured $w_\infty^{\Psi^{_+}}$, $w_L^{\Psi^{_+}}$, and $w_L^{\Psi^{_-}}$. The circles give the experimental data points and the lines are theoretical prediction obtained using average parameters obtained from the data. The vertical error bars were obtained by applying $\sqrt{N}$ fluctuations to the measured coincidence counts, then averaging over 100 iterations to obtain the standard deviation. The horizontal error bars are estimated to be $\pi/24$.}}
  \label{results}
\end{figure}

In figure \ref{results} we see the main result of our work: a single nonlinear witness is able to verify the entanglement of a large range of input states, that is states of the form either $\rho^{\Phi}$ or $\rho^{\Psi}$. In contrast, no single linear witness is able to verify entanglement over the same range; the expectation value of each linear witness is above zero for half of the states we measure.   Since we use quantum state tomography to confirm that our states are entangled, this means that each linear witness delivers an inconclusive result and thus cannot detect entanglement in a large range of states that are entangled.  These results are for the two-dimensional subspaces described in equations (\ref{6}) and (\ref{7}) where $\ell=2$.

In the anti-correlated case, all expectation values of the nonlinear witness are negative, indicating entangled states for all phases observed.  In the correlated case, the nonlinear witness is negative for the majority of the observed states. However, near $\pi/2$ and $3\pi/2$, there are three states that have slightly positive expectation values. This is attributable to noise introduced during measurement, resulting in lower purity of the state.
 

\vspace{0.5cm}
{\it Discussion}: Our results clearly show that our nonlinear witnesses are able to establish the entanglement of the relevant class of states. Nonetheless, it is interesting that the numerical values of the expectation value of the witness are quite different in the two cases, as the witnesses are defined in such a manner that we expect the values to be comparable.  Consequently, the difference between the correlated and anti-correlated cases highlights an interesting aspect of nonlinear entanglement witnesses: the extreme sensitivity of the expectation value with regards to the outcome of a single projective measurement.   As can be seen from equation (\ref{nlw}), the last term that is subtracted in the calculation is inversely proportional to $1 - |{\rm Tr}(\rho (-\sigma_z \otimes \sigma_z ))|^2$ (one minus the square of the contrast of the OAM correlations), and consequently, the precise expectation value that is measured is highly sensitive to ${\rm Tr}(\rho (-\sigma_z \otimes \sigma_z ))$.  As the strength of the OAM correlations depends critically on a few measurements, so too does the obtained value of $w_\infty$.   

The difference in the range of measured expectation values originates in the strength of the OAM correlations for each case.  For the anti-correlated states of the form $\rho^\Psi$, the average measured value of $|{\rm Tr}(\rho (-\sigma_z \otimes \sigma_z ))|$ was equal to 0.92, whereas for the correlated states of the form $\rho^{\Phi}$, the average measured value of $|{\rm Tr}(\rho (-\sigma_z \otimes \sigma_z ))|$ was equal to 0.69.   We attribute the reduced contrast in the correlated case to the asymmetry introduced by placing only one Dove prism in the system. The theoretical fits to the data are adjusted to reflect the appropriate measured contrasts.

In certain situations it would be desirable to increase further the range of states accessible to nonlinear witnesses. There are two possible avenues for doing so. The first method involves adjusting the exact form of the witness. The only degree of flexibility in the construction of the accessible nonlinear witnesses described in Ref.~\cite{Arrazola2012} is in the choice of unitary operator $U$. For our particular choice of $U$, a nonlinear improvement on a correlated linear witness will not be able to detect entanglement in anti-correlated states, and vice versa. Other choices of $U$ are possible, for example $U=(\mathbb{1}-\sigma_x \otimes \sigma_x-\sigma_y \otimes \sigma_y+\sigma_z \otimes \sigma_z)/2$; in this case, $U=2W_L^{\Psi^{_+}}$. With this choice of $U$, the nonlinear witness can access different states. More specifically, starting with a linear witness ($W_L^{\Psi^{_+}}$) that detects anti-correlated states, the nonlinear improvement using $U=2W_L^{\Psi^{_+}}$ can detect both anti-correlated and correlated states. 

The second method involves using two carefully chosen nonlinear witnesses. In fact, using two nonlinear witnesses, it is possible to extend the range sufficiently to detect entanglement in all qubit entangled states. For example, this can be achieved using the two witnesses shown in this paper: one witness that detects the correlated states (e.g.~$W_\infty^{\Phi^{_+}}$) and one witness that detects the anti-correlated states (e.g.~$W_\infty^{\Psi^{_+}}$). Using these two witnesses enables the verification of entanglement of the full range of pure quantum states, which includes all four Bell states, with only ten measurements. 


\vspace{0.5cm}
{\it Conclusions}: In this work we demonstrate experimentally that a single nonlinear witness is able to verify the entanglement of states of the form either $\rho^{\Phi}$ or $\rho^{\Psi}$; this is the largest range of entangled states detected with the fewest number of measurements.  This is a significant improvement over linear witnesses, which are not able to provide the same level of performance.  Such nonlinear witnesses require only a few measurements and no state reconstruction, thus drastically reducing processing time as compared to tomography. We also demonstrate laboratory procedures that allow us to vary the precise form of entangled quantum states, which provides an additional resource for quantum information protocols.   Thus, the combination of state preparation and nonlinear witnesses provides a clear indication of the significance of our approach to such applications as superdense coding and quantum teleportation.  Moreover, we envisage the continued application of nonlinear witnesses to other areas of quantum information science, where it is advantageous to extract maximal information with the minimum number of measurements.  


\vspace{0.5cm}
{\it Acknowledgements}: We thank J.~M.~Arrazola, O.~Gittsovich, and N.~L\"{u}tkenhaus for valuable discussions regarding this work. This work was supported by the Canada Excellence Research Chairs (CERC) Program and the Natural Sciences and Engineering Research Council of Canada (NSERC).

\vspace{0.5cm}
{\it Author Contributions}: The experiment was designed by J.~L.~and performed by J.~L., J.~Z.~S., and M.~A. The theoretical calculations and data analysis were performed by M.~A. All authors contributed to the writing of the manuscript.

\vspace{0.5cm}
{\it Methods}: We generate photon pairs entangled in the orbital angular momentum basis by means of parametric downconversion. A Nd:YAG laser at 355 nm with an average power of 150 mW is used to pump a 3-mm-long type I BBO crystal. We use a spatial light modulator (SLM) coupled with a single mode fibre in each arm of the system in order to select a particular mode of light. The SLMs display computer-generated holograms that modify the phase profile of the incoming light so that it is converted into the fundamental mode. The light in each arm then propagates to a single mode fibre, which creates an effective means of mode selection by only allowing the fundamental mode of light. The single mode fibres are connected to avalanche photodetectors and a coincidence counting card with a timing resolution of 25~ns.   The plane of the crystal is imaged onto the plane of the SLMs using a 4-f imaging system with a magnification of $\sim -3.33$; the focal lengths of the lenses are $f = 150$ mm and $f = 500$ mm. The planes of the SLMs are then imaged onto the fibre facets using a second 4-f imaging system with a magnification of $\sim -3.6 \times 10^{-3}$; the focal lengths of the lenses are $f = 400$ mm and $f = 1.45$ mm.

Each iteration of the experiment involves first preparing the state by setting the angle of each dove prism. Quantum state tomography is then performed on the photon pair to ensure that it is entangled and has the required phase. We use an overcomplete set of measurements in order to accurately determine the state, and we reconstruct the density matrix using the method in Ref.~\cite{Agnew2011}. We measure one nonlinear witness and two linear witnesses for each state. For the correlated case (one Dove prism), we measure the nonlinear witness $W_\infty^{\Phi^{_+}}$ and the linear witnesses $W_L^{\Phi^{_+}}$ and $W_L^{\Phi^{_-}}$. For the anti-correlated case (two Dove prisms), we measure the nonlinear witness $W_\infty^{\Psi^{_+}}$ and the linear witnesses $W_L^{\Psi^{_+}}$ and $W_L^{\Psi^{_-}}$. We repeat this process for varying values of $\varphi$ (correlated case) and $\phi$ (anti-correlated case) in order to produce the two plots in figure \ref{results}.

\bibliography{NonlinearWitness}
\bibstyle{unsrt}

\section{Supplementary Information}

\vspace{0.5cm}
{\it Linear witnesses}: For an anti-correlated state $\ket{\Psi^{_\pm}}$, the operator $W_L^{\Psi^{_\pm}}$ can be decomposed as follows:
\begin{align}
W_L^{\Psi^{_\pm}}&= \frac{1}{2} \Bigl( \ket{j, j}\bra{j, j}+\ket{k, k}\bra{k, k}  \Bigr) \notag\\
&\mp \frac{1}{2} \Bigl( \ket{x^{_+},x^{_+}}\bra{x^{_+},x^{_+}}+\ket{x^{_-},x^{_-}}\bra{x^{_-},x^{_-}}\notag\\
&-\ket{y^{_+},y^{_-}}\bra{y^{_+},y^{_-}}-\ket{y^{_-},y^{_+}}\bra{y^{_-},y^{_+}}   \Bigr).\label{decompositionanti}
\end{align}
In this case, since our OAM anti-correlations are of the form
\begin{align}
\ket{\Psi}&=\frac{1}{\sqrt{1+\varepsilon^2}} \Bigl( \ket{j, k}+\varepsilon e^{i\varphi}\ket{k, j}\Bigr)\notag\\
&=\frac{1}{\sqrt{1+\varepsilon^2}} \Bigl( \ket{0, 0}+\varepsilon e^{i\varphi}\ket{-\ell, \ell}\Bigr),
\end{align}
we note that $\ket{j}_A=\ket{0}$, $\ket{j}_B=\ket{\ell}$, $\ket{k}_A=\ket{-\ell}$ and $\ket{k}_B=\ket{0}$.

\vspace{0.5cm}
{\it How to generate a nonlinear witness}: A linear witness of a state $\rho$ with nonpositive partial transpose can be constructed using \cite{Lewenstein2000,Bruss2002}
\begin{equation}\label{mineigen}
W_L=(\ket{\eta}\bra{\eta})^{T_B},
\end{equation}
where $\rho^{T_B}$ denotes the partial transpose of $\rho$ on the Hilbert space of photon B and $\ket{\eta}$ denotes the eigenvector of $\rho^{T_B}$ corresponding to the minimum eigenvalue. Then an entangled state $\rho_{\rm ent}$ gives
\begin{equation}
w_1={\rm Tr}(\rho_{\rm ent}W_L)<0.
\end{equation}
In the following, we demonstrate how to generate the linear and nonlinear witnesses for the correlated Bell state
\begin{equation}\label{ent_corr2}
\ket{\Phi^{_+}}=\frac{1}{\sqrt{2}}\Bigl(\ket{j, j}+\ket{k, k}\Bigr).
\end{equation}
In this case, we find the eigenvector corresponding to the minimum eigenvalue of $\rho^{T_B} = (\ket{\Phi^{_+}}\bra{\Phi^{_+}})^{T_B}$ to be
\begin{equation}
\ket{\eta}=\frac{1}{\sqrt{2}} \Bigl(\ket{j, k}-\ket{k, j} \Bigr),
\end{equation}
which produces a linear witness
\begin{equation}
W^{\Phi^{_+}}_L=\frac{1}{2}\begin{pmatrix}
0 & 0 & 0 & -1 \\
0 & 1 & 0 & 0 \\
0 & 0 & 1 & 0 \\
-1 & 0 & 0 & 0 \end{pmatrix}.
\end{equation}
In the case of two qubits, the nonlinear improvement on the linear witness is \cite{Arrazola2012}
\begin{equation}
w^{\Phi^{_+}}_2={\rm Tr}(\rho W^{\Phi^{_+}}_L)-|{\rm Tr}(\rho(\rho_{\eta}U)^{T_B})|^2.
\end{equation}
We choose the operator $U$ to be $-\sigma_z\otimes\sigma_z$, which is equal to
\begin{equation}
U=\begin{pmatrix}
-1 & 0 & 0 & 0 \\
0 & 1 & 0 & 0 \\
0 & 0 & 1 & 0 \\
0 & 0 & 0 & -1 \end{pmatrix},
\end{equation}
and $\rho_{\eta} = \ket{\eta}\bra{\eta}$ is equal to
\begin{equation}
\rho_{\eta}=\frac{1}{2}\begin{pmatrix}
0 & 0 & 0 & 0 \\
0 & 1 & -1 & 0 \\
0 & -1 & 1 & 0 \\
0 & 0 & 0 & 0 \end{pmatrix}.
\end{equation}
This nonlinear witness can be iterated to further improve the strength of the witness \cite{Moroder2008}. When the number of iterations goes to infinity, the following witness is obtained \cite{Arrazola2012}:
\begin{align}
w^{\Phi^{_+}}_\infty(\rho)&={\rm Tr}(\rho W^{\Phi^{_+}}_L)-|{\rm Tr}(\rho(\rho_{\eta}U)^{T_B})|^2\notag\\
&-\frac{|{\rm Tr}(\rho \rho_{\eta}^{T_B})-{\rm Tr}(\rho(\rho_{\eta}U)^{T_B}){\rm Tr}(\rho U^{T_B})|^2}{1-|{\rm Tr}(\rho U^{T_B})|^2}.\label{infinity}
\end{align}
To calculate $w^{\Phi^{_+}}_\infty(\rho)$, we require knowledge of $W^{\Phi^{_+}}_L$, $U$ and $\rho_{\eta}$. However, the choice of the form of the unitary operator $U$ results in properties that minimise the number of required measurements for the nonlinear witness.  Since $U=-\sigma_z\otimes\sigma_z$, $\rho_{\eta}$ does not change when multiplied by $U$; it follows that
\begin{equation}
(\rho_{\eta}U)^{T_B}=\rho_{\eta}^{T_B}=W^{\Phi^{_+}}_L=\frac{1}{2}\begin{pmatrix}
0 & 0 & 0 & -1 \\
0 & 1 & 0 & 0 \\
0 & 0 & 1 & 0 \\
-1 & 0 & 0 & 0 \end{pmatrix}.
\end{equation}
This means that equation (\ref{infinity}) can be simplified to 
\begin{align}
w_\infty(\rho)&={\rm Tr}(\rho W^{\Phi^{_+}}_L)-|{\rm Tr}(\rho W^{\Phi^{_+}}_L)|^2 \\
&-\frac{|{\rm Tr}(\rho W^{\Phi^{_+}}_L)-{\rm Tr}(\rho W^{\Phi^{_+}}_L ){\rm Tr}(\rho U)|^2}{1-|{\rm Tr}(\rho U)|^2} \nonumber. 
\end{align}
We now see that the expectation values ${\rm Tr}(\rho W^{\Phi^{_+}}_L)$ and ${\rm Tr}(\rho U)$ are the only measurements that are required for the nonlinear witness.   The operator $W_L$, which is the standard linear witness, can be decomposed into six local measurements, and the operator $U$ requires four local measurements.   Two of the four measurements ($\ket{j, k}\bra{j, k}$ and $\ket{k, j}\bra{k, j} $) are in both $W^{\Phi^{_+}}_L$ and $U$; therefore, the nonlinear witness requires a total of eight measurements, which is approximately half the number required for tomography \cite{Thew2002}.

We note that for our choice of $U$, we obtain ${\rm Tr}(\rho U^{T_B})=-1$ for the maximally entangled state where $\varepsilon=1$, resulting in zero in the denominator. However, in reality, the state we detect is not pure; that is, the photons are in a state described by
\begin{equation}
\rho^{\psi}=\ket{\psi}\bra{\psi} p + \mathbb{1}(1-p)/4,
\end{equation}
where $p$ indicates the purity of the state and $\mathbb{1}$ represents the four-dimensional identity matrix. In this case, ${\rm Tr}(\rho^{\psi} U^{T_B})\neq-1$ for $p\neq1$, and equation (\ref{infinity}) is valid.

\end{document}